\documentstyle[12pt,amsfonts]{article}
\def\one{1\hskip-.37em 1}

\def\half{{\textstyle{\frac{1}{2}}}}

\def\ra{\rightarrow}

\def\>{\rangle}
\def\<{\langle}
\def\d{\delta}
\def\a{\alpha}
\def\g{\gamma}
\def\H{{\cal H}}
\def\E{{\mathbb E}\,}
\def\T{{\bf T}}
\def\D{{\cal D}}
\def\l{\lambda}
\def\tint{{\textstyle \int}}

\def\b{\begin{eqnarray*}}                  
\def\ee{\end{eqnarray*}}                    
\def\bn{\begin{eqnarray}}                  
\def\en{\end{eqnarray}}                    
\def\e{\epsilon}

\title{Universal Procedure for  \\
Enforcing Quantum Constraints}
\author{John R. Klauder\\
Departments of Physics and Mathematics\\
University of Florida\\
Gainesville, FL  32611}
\date{}                               
\begin{document}
\maketitle
\begin{abstract}
An abstract formulation of quantum dynamics in the presence of a 
general set of quantum constraints is developed. Our constructive 
procedure is such that the relevant projection operator onto the 
physical Hilbert space is obtained with a single, common integration 
procedure over the original Lagrange multiplier variables that is 
completely independent of the general nature of the constraints. In 
the associated lattice-limit formulation it is demonstrated that 
expansion of the constraint operator contribution to second order in 
the lattice spacing is necessary while, as usual, only a first-order 
expansion is needed for the dynamical operator contribution. Among 
various possibilities, coherent state path integrals are used to 
illustrate a completely functional representation of the abstract 
quantization procedure.
\end{abstract}
\section{Introduction}
The abstract operator formulation of quantum mechanics is generally 
recognized as being the most fundamental, while those formulations 
that involve one or another representation need to contend with 
representation-dependent issues (e.g., lack of global coordinates) 
that have no place in the essential formulation. Our goal here is to 
study quantum dynamics and quantum constraints as much as possible 
in an abstract, representation-independent way. Only then should one 
feel comfortable about introducing a representation for further 
analysis. Let us start with some basics.
 
The evolution operator, from time $t=0$ to time $t=T>0$, for a general 
quantum system with an abstract, self-adjoint, time-independent 
Hamiltonian operator $\H$, is given, in units where $\hbar=1$ (which 
are often used), by
the unitary operator $e^{-i\H T}$ acting on an abstract 
Hilbert space $\frak H$ known henceforth as the ``original'' Hilbert 
space. If the Hamiltonian is time dependent and self adjoint for all 
$t$, $0\le t\le T$, the evolution operator is instead given by the 
unitary operator
  $$ {\T}\,e^{-i\tint_0^T\,\H(t)\,dt}\;,  $$ where $\T$ 
denotes the time-ordering operator. We may 
append to $\H(t)$ a sum of additional, time-independent, self-adjoint 
operators $\{\Phi_\a\}_{\a=1}^A$ with general, real-valued, time-dependent 
coefficients $\{\l^\a(t)\}_{\a=1}^A$ leading, by assumption, for all $t$, 
$0\le t\le T$, to the self-adjoint operator 
  $\H(t)+\l^\a(t)\Phi_\a$, with summation implied (the generalization 
to a set of time-dependent, self-adjoint operators 
$\{\Phi_\a(t)\}_{\a=1}^A$ is briefly discussed later). Taken as 
the generator of time translations, the extended evolution operator reads
  $$ {\T}\,e^{-i\tint_0^T[\H(t)+\l^\a(t)\Phi_\a]\,dt}\;. $$
If the functions $\{\l^\a(t)\}_{\a=1}^A$ are fixed from the beginning 
and held fixed, the term $\l^\a(t)\Phi_\a$ provides an addition to the 
Hamiltonian and modifies the dynamics in the expected way. If we regard 
the functions $\{\l^\a(t)\}_{\a=1}^A$ as variable external sources, 
then functional derivatives with respect to them lead to prototype 
expressions from which time-ordered correlation functions may be 
obtained. Alternatively, if the functions $\{\l^\a(t)\}_{\a=1}^A$ 
denote suitable stochastic variables, then an expectation of the
given expression may relate to systems interacting with random
potentials.
Besides these examples, let us consider a rather different  
interpretation and application of the extra term $\l^\a(t)\Phi_\a$, 
which we now describe.  

Let us accept the hypothesis that the set $\{\Phi_\a\}_{\a=1}^A$ 
consists of {\it constraint operators}, which, when appropriately satisfied, 
direct attention to a subspace of $\frak H$ called the ``physical'' 
Hilbert space and which is denoted by ${\frak H}_{\,\rm phys}$. As a 
closed subspace, it follows that ${\frak H}_{\,\rm phys}=\E{\frak H}$ 
for some unique projection operator $\E\,(=\E^\dagger=\E^2)$. In turn, 
the projection operator $\E$ should be given by an acceptable function 
of the constraint operators $\{\Phi_\a\}^A_{\a=1}$, and it is the role 
of an integration over the variables $\{\l^\a(t)\}_{\a=1}^A$ to build 
the necessary projection operator $\E$. Hereafter, we shall refer to the 
set of functions $\{\l^\a(t)\}_{\a=1}^A$ as the {\it Lagrange multiplier 
variables}.
   
In simple cases it is easy to determine a suitable projection operator 
$\E$ \cite{kla}. For example, if $A=1$, and so there exists a single 
constraint, then---see Appendix A---we observe, for any 
$\d=\d(\hbar)>0$ $(\!\!($N.B. $\d(\hbar)$ is {\it not} a Dirac delta 
function but a rather general function of $\hbar$$)\!\!)$, that 
$$ \int_{-\infty}^\infty\,e^{-i\l\Phi}\,\frac{\sin(\delta\l)}{\pi\l}\,d\l
=\E(-\d\le\Phi\le \d)=\E(\Phi^2\le \d^2)\;. $$
If $\Phi=0$ is part of a purely discrete spectrum near zero, then it 
is possible to choose $\d>0$ such that $\E(\Phi^2\le\d^2)=\E(\Phi=0)$. 
When $\Phi=0$ is part of the continuum, we choose $0<\d\ll 1$; one 
procedure to take the limit $\d\ra0$ to arrive at the true physical 
Hilbert space ${\frak H}_{\,\rm true\;phys}$ is discussed in 
Appendix B. From this discussion, it follows in the continuum case that 
we may 
regard ${\frak H}_{\,\rm phys}$ as a provisional physical Hilbert space, 
which, despite the qualifying adjective, is, for an extremely small $\d$, 
already acceptable for calculational purposes.
As a second example consider that the constraints are the generators 
of a compact Lie group, and let $dm(\l)$ denote the normalized group 
invariant measure, $\tint dm(\l)=1$. In that case (e.g., \cite{kla,sha,gov})
  $$ \int e^{-i\l^\a\Phi_\a}\,dm(\l)=\E(\Phi_\a=0, 1\le\a\le A)=
\E( \Phi_\a\d^{\a\beta}\Phi_\beta=0)\;. $$

In order to generate the projection operator $\E$, these examples 
illustrate that different measures for the Lagrange multiplier variables 
may be involved. Indeed, in order to construct $\E$, even for a single 
set of constraints, it is generally the case that the measure for the 
Lagrange multiplier variables is not unique.

The use of different measures for different situations is not wrong but 
it does mean that some specific properties about the system, especially 
the constraints, need to be determined before an expression for $\E$ can 
be found. This is clearly a nuisance, and it raises the question whether 
there exists a single, common, integration procedure by which we can 
construct $\E$, and which, moreover, is a procedure that is {\it completely 
independent of the specific constraints at hand}. The answer to this 
question is ``Yes!'', and the aim of the present paper is to develop 
a {\it universal prescription for determining $\E$} from a given set of 
constraint operators, whatever kind of constraints they may be, by means 
of a prescription that is also independent of the specific Hamiltonian 
operator.

There are various kinds of constraints that need to be considered. In the 
usual nomenclature (see, e.g., \cite{hen}), the constraints may be: 
first class, closed or open; anomalous or second class; primary or 
secondary; irreducible or reducible; regular or irregular. In the 
reducible case, where the constraints are linearly dependent, it is 
possible that there is an infinite degeneracy when $A=\infty$ even if 
the number of degrees of freedom is finite; see below for an example. 
Regularity refers to whether the classical constraint is taken (say) as 
$q=0$ or as $q^3=0$, etc., along with the quantum repercussions of this 
kind of multiple choice test. Implicitly, we assume that the Hamiltonian 
and the constraints satisfy some minimal consistency condition, in 
particular, that the set of constraints already includes all the necessary 
constraints for a given Hamiltonian.. 

In what follows, all varieties of constraints will be dealt with in a 
uniform fashion.

\section{Construction of the Physical Space\\ Projection and Evolution 
Operators}
We first observe that the extended evolution operator may be written in the 
form of a lattice limit given by
$$ \lim_{\e\ra0}\prod_{1\le n\le N}^{\longleftarrow} \bigg[
\T e^{-i\tint_{(n-1)\e}^{n\e}\,\H(t)\,dt}\bigg]\,\bigg[\T 
e^{-i\tint_{(n-1)\e}^{n\e}\,\l^\a(t)\Phi_\a\,dt}\bigg]\;, $$
where $\e\equiv T/N$ and the directed product (symbol $\longleftarrow$) 
also respects the time ordering. Thus, this expression is simply an 
alternating sequence of short-time evolutions, first by $\l^\a(t)\Phi_\a$, 
second by $\H(t)$, a pattern which is then repeated $N-1$ more times. The 
validity of this Trotter-product form follows whenever 
$\H(t)^2+\Phi_\a\d^{\a\beta}\Phi_\beta$ is essentially self adjoint 
for all $t$, $0\le t\le T$. As a slight generalization, we shall assume 
that $\H(t)^2+\Phi_\a M^{\a\beta}\Phi_\beta$ is essentially self adjoint 
for all $t$, $0\le t\le T$. Here the real, symmetric coefficients 
$M^{\a\beta}\,(=M^{\beta\a})$ are the elements of a positive-definite 
matrix, i.e., $\{M^{\a\beta}\}>0$. For a finite number of constraints, 
$A<\infty$, it is sufficient to assume that $M^{\a\beta}=\delta^{\a\beta}$; 
other choices for $M^{\a\beta}$ may be relevant when $A=\infty$. 

With all this in mind, we shall explain the construction of a formal 
integration procedure whereby
 $$ \int \T e^{-i\tint_{(n-1)\e}^{n\e}\,\l^\a(t)\Phi_\a\,dt}\,\D R(\l)=
\E(\!\!(\Phi_\a M^{\a\beta}\Phi_\beta\le \d(\hbar)^2)\!\!)\;, $$
and for which the integral represented by $\tint\cdots\D R(\l)$ is 
independent of the set of operators $\{\Phi_\a\}$ and the Hamiltonian 
operator $\H(t)$ for all $t$. We will define the left-hand side of this 
expression by means of an additional lattice limit. 

\subsection*{Constructing the basic projection operator}
For any given $n$, $1\le n\le N$, let us divide the time interval 
$((n-1)\e,n\e]$ into 
$N'$, $N'<\infty$, steps of equal duration, $\e'\equiv\e/N'=T/(NN')>0$. 
In this way we are led to consider the integrated directed product
  $$\int \prod_{1\le l\le N'}^{\longleftarrow}\,e^{-i\e'\l_{r(n,l)}^\a
\Phi_\a}\,dR(\l_{r(n,l)}) $$
for some measure $dR(\l_{r(n,l)})$ to be determined.
Observe that in the indicated expression the continuum variable $t$ has 
been replaced by steps of size $\e'$ consecutively numbered by $r(n,l)
\equiv (n-1)N'+(l-1)$ for $1\le n\le N$ and $1\le l\le N'$. As far as 
the previous integral goes, the particular value of $n$ is unimportant 
and we will suppress it for the rest of this discussion.  Thus for the 
present we replace $r(n,l)$ by $l$ for simplicity. Let us proceed in 
small steps.

Introduce a measure $dS(\l_l)$ for each $\l_l$ and an integral over 
$\l=\{\l_l\}_{l=1}^{N'}$ given by
\b && J\equiv\int\prod_{1\le l\le N'}^{\longleftarrow}\,e^{-i\e'\,\l_l^\a
\Phi_\a}\,dS(\l_l)\\
  &&\hskip.35cm=\int \prod_{1\le l\le N'}^{\longleftarrow}\,[\!\![\,
\one-
i\e' \,\l_l^\a\Phi_\a-\half\e'^{\,2}\,\l_l^\a \l^\beta_l \Phi_\a
\Phi_\beta +O(\e'^{\,3})\,]\!\!]\;dS(\l_l)\;. \ee
We choose $dS(\l_l)$ as a normalized probability measure with the first 
two correlation functions (moments) given by $\tint \l^\a_l\,dS(\l_l)
\equiv 0$ and $\tint \l^\a_l\l^\beta_l\,dS(\l_l)\equiv (2\g/\e')\,
M^{\a\beta}$, $\g>0$, values that are independent of the time-slice 
label $l$. With a minimal requirement on higher moments (such as finite 
fourth-order moments), it follows that
\b &&J=\prod_{1\le l\le N'}^{\longleftarrow}\,[\!\![\one-\e'\g\,(\Phi_\a 
M^{\a\beta}\Phi_\beta)]\!\!] +O(\sqrt{\e'})\\
 &&\hskip.4cm =\,e^{-N'\e'\g\,(\Phi_\a M^{\a\beta}\Phi_\beta)}\,+
O(\sqrt{\e'})\;, \ee
where the last expression follows because the factors in the directed 
product are now identical
for every time slice $l$.
At this point we may take the limit $N'\ra\infty$, $\e'\ra0$ with 
$N'\e'=\e$ held fixed. Stated formally, the net result is that
$$\int\T e^{-i\tint_{(n-1)\e}^{n\e}\,\l^\a(t)\Phi_\a\,dt}\,\D S(\l)=
e^{-\e\g(\Phi_\a M^{\a\beta}\Phi_\beta)}\;, $$
independently of $n$, $1\le n\le N$.
Note that many choices of the basic measure $dS(\l_l)$ will lead to 
the same result. The mechanism which lies behind this conclusion is 
similar to one leading to the central limit theorem in probability.
 
The next step in the process is to analytically extend the variable 
$\g$, $\g\ra\g+i\tau$, where $\g>0$ still and $-\infty<\tau<\infty$, 
namely 
$$ e^{-\e\g(\Phi_\a M^{\a\beta}\Phi_\beta)}\ra e^{-\e(\g+i\tau)(
\Phi_\a M^{\a\beta}\Phi_\beta)}\;, $$
and after achieving this extension, the following step is to pass to the 
limit $\g\ra0^+$. Both of these steps are achieved by the single 
integral operation
 \b &&\int e^{-\e\g(\Phi_\a M^{\a\beta}\Phi_\beta)}\,d\Gamma(\g)\\
&&\hskip1cm\equiv\lim_{B\ra\infty}\,\sqrt{B/2\pi}\,
e^{\frac{1}{2} B\tau^2}\int_0^\infty e^{iB\g\tau-\frac{1}{2} B\g^2}\, 
e^{-\e\g(\Phi_\a M^{\a\beta}\Phi_\beta)}\,d\g\\
  &&\hskip1cm =e^{-i\e\tau(\Phi_\a M^{\a\beta}\Phi_\beta)}\;. \ee
The final step in the construction---see Appendix A---involves an 
integration over $\tau$ given by
  \b && \hskip-1cm\int e^{-i\e\tau(\Phi_\a M^{\a\beta}\Phi_\beta)}
\,d{\sf T}(\tau)\\ 
&& \hskip.5cm\equiv\lim_{\zeta\ra0^+}\lim_{L\ra\infty}\,
\int_{-L}^L\,e^{-i\e\tau(\Phi_\a M^{\a\beta}\Phi_\beta)}\,
\frac{\sin[\e(\d^2+\zeta)\tau]}{\pi\tau}\,d\tau\\
&&\hskip.5cm=\E(\e\,\Phi_\a M^{\a\beta}\Phi_\beta\le\e\,\d^2)\\
 &&\hskip.5cm=\E(\Phi_\a M^{\a\beta}\Phi_\beta\le\d^2)\;, \ee
which achieves our goal. We occasionally symbolize the collective 
set of operations by $\tint\cdots\D R(\l)$, leaving the integrals 
over $\g$ and $\tau$ implicit. 
 
To summarize, let us assemble the several, separate operations together. 
In particular,  we observe that \b &&\hskip-1.3cm\E(\Phi_\a 
M^{\a\beta}\Phi_\beta\le\d^2)=\int \D R(\l)\;\T \,
e^{-i\tint_{(n-1)\e}^{n\e}\l^\a(t)\Phi_\a\,dt}\\
&&\hskip2.29cm\equiv\lim_{\zeta\ra0^+}\lim_{L\ra\infty}
\int_{-L}^L\frac{\sin[\e(\d^2+\zeta)\tau]}{\pi\tau}\,d\tau\;\\
&&\hskip3.3cm\times\lim_{B\ra\infty}\sqrt{B/2\pi}\,e^{\frac{1}{2} 
B\tau^2}\int_0^\infty e^{iB\g\tau-\frac{1}{2} B\g^2}\,d\g\\
&&\hskip3.3cm\times\int\,\D S(\l)\;\T\, 
e^{-i\tint_{(n-1)\e}^{n\e}\l^\a(t)\Phi_\a\,dt}\;, \ee
where the proper definition of $\tint\cdots\D S(\l)$ has been given 
above. In accord with the discussion in Appendix A, we note that if 
the final limit is replaced by $\lim_{\zeta\ra0^-}$, the result 
becomes $\E(\Phi_\a M^{\a\beta}\Phi_\beta<\d^2)$.

A few comments are in order. It is noteworthy that $\tint \,\D R(\l)=1$. 
Thus, in the hypothetical case that all the constraint operators are 
bogus, i.e., $\Phi_\a\equiv0$ for all $\a$ (i.e., not simply 
{\it weakly} zero but {\it strongly} zero), it would follow that 
 $$  \int \one\,\D R(\l)=\E(0\le\d^2)\equiv\one\;.  $$
If some, but not all, of the constraint operators are bogus, then, 
even though an integration over the associated Lagrange multiplier 
variables takes place, the answer will be identical to one in which 
the Lagrange multiplier variables for the bogus constraint operators 
were ignored from the very beginning. As another hypothetical case, 
assume that there is an infinite number of identical constraint 
operators, for example, $\Phi_\a\equiv\Phi$ for all $\a$, 
$\a=1,2,3,\dots\,$. Consequently, the extended evolution operator 
involves  $\H(t)+\l^\a(t)\Phi_\a= \H(t)+\Sigma_{\a=1}^\infty\l^\a(t)\,\Phi$. 
We clearly see that there is only one relevant Lagrange multiplier 
variable, but to act on that knowledge would amount to choosing an 
irreducible set of constraints from the original set. Instead, let us 
keep all the Lagrange multiplier variables as independent quantities 
and proceed along the lines developed in this paper. In this case let 
us choose, for example,  $M^{\a\beta}\equiv(6/\pi^2\a^2)\,\d^{\a\beta}$. 
Then it follows that the result of the integration with respect to 
$\D R(\l)$ yields the projection operator
$$  \E(\Phi_\a M^{\a\beta}\Phi_\beta\le\d^2)=\E(\!\!(
\Sigma_{\a=1}^\infty\,(6/\pi^2\a^2)\,\Phi^2\le\d^2)\!\!)=
\E(\Phi^2\le\d^2)\;, $$
as desired. This is an (extreme) example of how the present procedure 
deals with a reducible set of 
constraints without needing to extract an irreducible subset. For more 
complicated examples it may be useful to experiment with different 
choices of $M^{\a\beta}$.

\subsection*{The full-time physical evolution operator}
Up to this point we have determined how to construct the desired 
projection operator $\E$ for any (nonzero) short-time interval $\e=T/N$. 
Let us assume that we have carried out the necessary Lagrange multiplier 
variable integrations for every time slice, i.e., all $n$, $1\le n\le N$. 
The result will be the evolution operator restricted to the physical 
Hilbert space---let us call it ${\cal E}(T)$---which is given by
 $${\cal E}(T)\equiv\lim_{\e\ra0}\,\T 
e^{-i\tint_{(N-1)\e}^{N\e}\H(t)\,dt}\,\E\,
e^{-i\tint_{(N-2)\e}^{(N-1)\e}\H(t)\,dt}\,\E\,
\cdots\,\E\,e^{-i\tint_{0}^{\e}\H(t)\,dt}\,\E\;. $$
The indicated limit can be taken with the result that
 $${\cal E}(T)=\T\,\E \,e^{-i\tint_0^T\,(\E\H(t)\E\!)\,dt}\,\E=
\E\,[\T \,e^{-i\tint_0^T\,(\E\H(t)\E\!)\,dt}\,]\,\E\;. $$
This result holds whether or not the effective Hamiltonian $\H_\E(t)
\equiv\E\H(t)\E$ at time $t$ is (essentially) self adjoint or merely 
symmetric (Hermitian) on ${\frak H}_{\,\rm phys}$. If it is not self 
adjoint, the resulting evolution will not be unitary. On the other hand, 
there are many cases of interest when $\E\H(t)\E$ will be self adjoint 
for all $t$, $0\le t\le T$, and the evolution on the physical Hilbert 
space will be by unitary operators $\cal E$(T) acting on 
${\frak H}_{\,\rm phys}$, and which, in fact, act as the zero operator 
on the Hilbert space ${\frak H}_{\,\rm phys}^{\,\perp}={\frak H}
\ominus{\frak H}_{\,\rm phys}$, namely, the subspace othogonal to 
${\frak H}_{\,\rm phys}$, which may be called the ``unphysical'' 
Hilbert space ${\frak H}_{\,\rm unphys}={\frak H}^{\,\perp}_{\,\rm phys}$. 
This follows from the fact that $\E{\frak H}_{\,\rm phys}^{\,\perp}\equiv0$. 
For convenience, we will assume that the effective Hamiltonian is self 
adjoint and therefore that the resultant evolution is unitary. 
$(\!\!($Note, even when the effective Hamiltonian is not self adjoint, 
that fact alone does not imply that the resultant evolution takes physical 
vectors 
into unphysical vectors; whatever the form of evolution, vectors that 
start in 
the physical Hilbert space will remain in the physical Hilbert 
space.$)\!\!)$

\subsubsection*{Constructive approach}
It is very instructive to see the lattice-limit construction for the 
evolution operator combined with the constraint operators in as direct 
a form as possible. For this purpose we assume that $\H(t)$ is a 
continuous function of $t$. In that case, and modulo domain considerations, 
we assert that
\b &&\hskip-.2cm\E\,[\T\,e^{-i\tint_0^T(\E\H(t)\E\!)\,dt}]\,\E \\
 &&=\lim_{\e\ra0}\int\prod_{1\le n\le N}^{\longleftarrow}\,
d{\sf T}(\tau_n)\,d\Gamma(\g_n)\,\lim_{\e'\ra0}\int
\prod_{1\le l\le N'}^{{\longleftarrow}} dS(\l_{r(n,l)})\\
&&\hskip2cm\,\times[\!\![\one -i\e'\H_{r(n,l)}-i\e'\l_{r(n,l)}^\a\Phi_\a
- -\half\e'^{\,2}\l_{r(n,l)}^\a\l_{r(n,l)}^\beta\Phi_\a\Phi_\beta+\,
\cdots\,]\!\!]\;. \ee
Here the single index $n$ on the variables $\tau$ and $\gamma$ labels 
consecutive time steps with a spacing $\e$.
Observe that only the {\it first-order term} (i.e., proportional to 
$\e'$) is necessary to include for the Hamiltonian, but to properly 
include the constraints it is necessary to include {\it first- and 
second-order terms} (i.e., proportional to $\e'$ and  $\e'^{\,2}$). 
This difference between the dynamics and the constraints holds because 
the integration over the Lagrange multiplier variables is, in reality, 
an integral over a (regulated) $\d$-correlated distribution.

This formula may also be brought to yet another interesting form. By 
carrying out the integrations over all $\l_{r(n,l)}$ and $\g_n$ we are 
left with the expression
\b &&\hskip-1cm\E\,[\T\,e^{-i\tint_0^T(\E\H(t)\E\!)\,dt}]\,\E \\
 &&=\lim_{\e\ra0}\int\prod^{\longleftarrow}_{1\le n\le N}\,
e^{-i\e\H_n-i\e\tau_n(\Phi_\a M^{\a\beta}\Phi_\beta)}\,d{\sf T}(\tau_n)\;, 
 \ee
where $\H_n=\H_{r(n,1)}$ and 
$$ \int(\cdots)\, d{\sf T}(\tau_n)\equiv\lim_{\zeta\ra0^+}
\lim_{L\ra\infty}\int_{-L}^L(\cdots)\,\frac{\sin[\e(\d^2+\zeta)\tau_n]}
{\pi\tau_n}\,d\tau_n \;. $$
This expression may be a good place to start calculations in certain 
circumstances.

\subsubsection*{Special case}
A great deal of simplification takes place whenever $\H(t)\E=\E\H(t)$ 
for all $t$, $0\le t\le T$. In this case we have the identity
$$\E\,[\T\,e^{-i\tint_0^T(\E\H(t)\E\!)\,dt}\,]\,\E=[\T\,
e^{-i\tint_0^T\H(t)\,dt}\,]\,\E\;.$$
Observe that the evolution is then automatically unitary in the 
physical Hilbert space, and that it is only necessary to impose the 
projection operator $\E$ once, say, as we have done, at the beginning 
of the evolution, very much in the manner of an initial-value equation 
used in the case of classical first-class constraints. 
\subsubsection*{Remark}
We hasten to add that for a given, specific set of constraints it is 
generally possible to construct $\E$ by alternative and possibly 
simpler integral representations. By all means, when simpler procedures 
exist they should be used. For example, as noted earlier in the case of 
a set of constraint operators which close to form a 
Lie algebra for a compact group, 
a group invariant average over all unitary transformations continuously 
connected to the identity and generated by the constraint operators leads 
to the same projection operator $\E$ given by the general prescription 
developed in this paper for sufficiently small $\d$, $\d>0$. 

\subsection*{Relation with Dirac's prescription}
The machinery developed above is expressed  entirely in an abstract 
operator form plus a functional integral over the Lagrange multiplier 
variables which has a well-defined formulation as a lattice limit. 
Another abstract approach, and one of the more common methods for 
dealing with constraints in the presence of dynamics, is that proposed 
by Dirac \cite{dir}. Briefly summarized, Dirac's procedure defines the 
physical Hilbert space as the subspace spanned by vectors 
$|\psi\>_{\,\rm phys}$ that fulfill the requirement  
$$ \Phi_\a\,|\psi\>_{\,\rm phys}=0$$
for all $\a$, implying an exact fulfillment of the quantum constraints. 
For second-class constraints, such as $\Phi_1=P$ and $\Phi_2=Q$, where 
$P$ and $Q$ are conventional Heisenberg variables, there is no nontrivial 
solution of the equations $P|?\>=Q|?\>=0$ for that would require 
$[Q,P]|?\>=i\hbar|?\>=0$.  Moreover, for a constraint $\Phi$ with zero 
in its continuous spectrum, the equation $\Phi|\#\>=0$ has no 
(normalizable) solution in the Hilbert space, i.e., there is no 
nontrivial solution for $|\#\>$. Thus, for second-class constraints, and 
for general, open first-class constraints, as well as for constraints 
with zero in the continuum, the basic Dirac prescription is unsatisfactory. 
Of course, there are modifications of the Dirac scheme, such as the use 
of a Dirac bracket \cite{dir} in the classical theory followed by the 
adoption of this bracket in the process of canonical quantization. 
Unfortunately, the classical phase space associated with the Dirac 
bracket generally has a nontrivial curvature (and possibly a nontrivial 
topology), and so a straightforward application of the quantization 
rules generally leads to an incorrect answer.

Nevertheless, there are some elements of the Dirac procedure and the 
procedure developed in the present paper that are rather close to each 
other, at least in spirit. For one thing, we note that if $\Phi_\a|?\>=0$ 
for all $\a$, then $(\Phi_\a \d^{\a\beta}\Phi_\beta)|?\>=0$ and 
conversely, where summation is implied and we are only considering the 
case $M^{\a\beta}=\d^{\a\beta}$ for simplicity. For the second-class 
constraints considered above we would be asking that $(P^2+Q^2)|?\>=0$, 
which, as is well known, implies that $|?\>\equiv0$, an unsatisfactory 
solution. Let us relax this condition (cf., \cite{mcm}) and say that 
$(P^2+Q^2)|?\>=\hbar\,|?\>$, for which we do have a nonvanishing solution, 
namely $|?\>=|0\>$, the ground state of the harmonic oscillator. There is 
no violation of the classical limit for these constraints by this 
modification since, as $\hbar\ra0$, we recover the original, classical, 
second-class constraints, namely, $p=0$ and $q=0$, in a standard notation. 
However, the generalization of the Dirac procedure just illustrated 
suffers from one important defect:   
The extended criterion does not satisfactorily specify the physical 
Hilbert space. Take the given example and instead let $(P^2+Q^2)|?\>=
c\hbar|?\>$ for some constant $c$. If $c=3$ there is a solution which 
is $|?\>=|1\>$, the first excited state, but for $c=2$---or indeed for 
any $c$, $1< c< 3$---there is {\it no} nontrivial solution and so $|?\>=0$, 
even though for any of these cases the recovery of the classical 
second-class constraints still applies.

Rather than relax the original Dirac prescription in the manner just 
discused, we have chosen to look for solutions to the equation 
$\E(\!\!(\Phi_\a\d^{\a\beta}\Phi_\beta\le\d^2(\hbar))\!\!)|?\>=|?\>$. 
It is convenient to examine this expression as a function of $\d$. For 
sufficiently large $\d$, there is generally a nontrivial subspace of 
vectors satisfying this relation. As $\d$ decreases, the space in 
question remains the same or its span may decrease. As an example, 
again take the second-class case and consider 
$\E(P^2+Q^2\le c\hbar)|?\>=|?\>$. 
Beginning with small $c$, we observe for $c<1$ that there is no 
nontrivial solution, $|?\>=0$; for  $ 1\le c<3$, there is a single 
nontrivial solution, $|?\>=|0\>$, which spans a one-dimensional 
Hilbert space; while for $3\le c<5$, there is a two-dimensional 
Hilbert space spanned by the two vectors $|0\>$ and $|1\>$, where the 
latter vector is the first  excited state of the harmonic oscillator; 
etc. This dependence on the parameter $c$ is to be preferred to that 
described above.

Suppose the second-class constraints are irregular as well. In particular, 
let $\Phi_1=P$ and $\Phi_2=Q^3$. We are then led to consider 
$\E(P^2+Q^6\le c'\hbar^{3/2})$. Let $c'=c'_0$ denote the value that 
captures the ground state of the ``Hamiltonian'' $P^2+Q^6$, let 
$c'=c'_1$ denote the value that captures the first excited state, etc. 
The argument then proceeds as above. Of course,
$\E(P^2+Q^6\le c'\hbar^{3/2})\not=\E(P^2+Q^2\le c\hbar)$, for general 
$c$ and $c'$, but this fact merely represents the ever-present 
$O(\hbar)$ ambiguity in any quantization procedure. 

\subsection*{Time-dependent constraints}
In the time-dependent case $\Phi_\a\ra\Phi_\a(t)$, for all $\a$ and 
$t$, $0\le t\le T$, and assuming each constraint operator is continuous 
in $t$ as well as the needed self-adjoint properties, we are quickly led 
to an expression for the evolution operator in the {\it time-dependent} 
physical Hilbert space, given by
\b &&\hskip-.2cm{\cal E}'(T) \\
&&\hskip-.2cm\equiv\lim_{\e\ra0}\,\T e^{-i\tint_{(N-1)\e}^{N\e}\H(t)\,dt}
\,\E_{N-1}\,e^{-i\tint_{(N-2)\e}^{(N-1)\e}\H(t)\,dt}\,\E_{N-2}\,\cdots
\,\E_1\,e^{-i\tint_{0}^{\e}\H(t)\,dt}\,\E_0\,. \ee 
Here 
 \b &&\E_n\equiv\E(\!\!(\Phi_\a(n\e) M^{\a\beta}\Phi_\beta(n\e)
\le\d^2)\!\!)\\
   &&\hskip.69cm =\int\T e^{-i\tint_{n\e}^{(n+1)\e}\l^\a(t)\Phi_\a(t)
\,dt}\,\D R(\l)\;. \ee
While this expression is correct, there does not seem to be any simpler 
and more elegant operator expression that holds in the general case of 
time-dependent constraints. 

We do not pursue time-dependent constraints further, and hereafter we 
again assume that all the constraints have no explicit time dependence.

\section{Functional Integral Representation}
It is convenient to adopt a functional representation for the original 
Hilbert space $\frak H$, and to present the main properties of the 
foregoing argument in that language. For this purpose, let us focus on a 
canonical system 
with ${\rm dim}({\frak H})=\infty$ and $J$ degrees of freedom, 
$J<\infty$,  described by the basic, irreducible, self-adjoint 
operators $P=\{P^j\}$and $Q=\{Q^j\}$, $1\le j\le J$, which satisfy the 
canonical commutation relations $[Q^j,\,P^k]=i\d^{jk}\one$ with all other 
commutation relations vanishing. As an irreducible set of operators, 
all other operators may be constructed from the basic ones. Hence, 
$\H(t)=\H(P,Q,t)$ and $\Phi_\a=\Phi_\a(P,Q)$ for some suitable functions. 

Although other bases may be used, we concentrate here on the 
functional representation induced by coherent states.
Thus, along with the canonical operators, we introduce canonical 
coherent states defined \cite{kls} by
  $$|p,q\>\equiv e^{-iq\cdot P}\,e^{ip\cdot Q}\,|\eta\>\;, $$
where $p=\{p^j\}^J_{j=1}$ and $q=\{q^j\}^J_{j=1}$, with each $p^j$ and 
$q^j$ an 
arbitrary real number, $q\!\cdot\!  P=\Sigma_{j=1}^J\,q^jP^j$, etc., 
and $|\eta\>$ is a fixed, normalized fiducial vector. Moreover, for any 
choice of $|\eta\>$ it follows \cite{mck} that
  $$\int |p,q\>\<p,q|\,d\mu(p,q)=\one\;,\hskip1.2cm d\mu(p,q)=
\Pi_{j=1}^J\,dp^j\,dq^j/2\pi\;. $$
Needed domain restrictions on $|\eta\>$ will be assumed, without comment, 
as they arise.
At this point $p$ and $q$, although having suggestive names, are just 
mathematical labels of the coherent states. As a first step in giving 
meaning to $p$ and $q$, 
we assume that $|\eta\>$ is ``physically centered'', meaning that 
$\<\eta|P|\eta\>=0$ and $\<\eta|Q|\eta\>=0$. In this case it follows that 
$\<p,q|P|p,q\>=p$ and $\<p,q|Q|p,q\>=q$. As a second step, we assume 
that $|\eta\>$ is ``physically scaled'', i.e., that $\<\eta|P^2|\eta\>=
O(\hbar^a)$ and $\<\eta|Q^2|\eta\>=O(\hbar^b)$, where $a>0$, $b>0$, 
and $a+b=2$.

To obtain the desired functional representation for the lattice 
formulation we repeatedly insert coherent-state resolutions of unity 
leading to 
\b &&\hskip-.3cm\<p'',q''|\,\E[\,\T\,e^{-i\tint_0^T(\E\H(t)\E\!)\,dt}
\,]\E|p',q'\>\\
 &&\hskip0cm=\lim_{\e\ra0}\int\!\prod_{n=1}^Nd{\sf T}(\tau_n)\,
d\Gamma(\g_n)\bigg\{\,\lim_{\e'\ra0}\int\!\prod_{l=1+\d_{n1}}^{N'}
\bigg[ d\mu(p_{r(n,l)},q_{r(n,l)})\,dS(\l_{r(n,l)})\bigg]\\
&&\hskip1.4cm\times\prod_{l=1}^{N'}\bigg[\,\<p_{r(n,l)+1},q_{r(n,l)+1}|
\,[\!\![\one
- -i\e'\H_{r(n,l)}\\
&&\hskip3cm-i\e'\l_{r(n,l)}^\a\Phi_\a-
\half\e'^{\,2}\l_{r(n,l)}^\a\l_{r(n,l)}^\beta\Phi_\a\Phi_\beta]\!\!]
\,|p_{r(n,l)},q_{r(n,l)}\>\,\bigg]\bigg\}\;. \ee
In this expression we have again used $r(n,l)\equiv (n-1)N'+(l-1)$ and 
identified $(p'',q'')=(p_{NN'},q_{NN'})$ and $(p',q')=(p_{0},q_{0})$. 
We reemphasize that it is necessary to expand the constraint operators 
to second order in the lattice spacing in order to  correctly obtain the 
desired projection operator. 

It is customary to interchange the order of integration and the continuum 
limit in order to arrive at a formal expression for the propagator in the 
presence of the constraints. In so doing one adopts for the integrand the 
expression it would assume for continuous and differential paths (even if 
such paths do not contribute in the continuum limit!). In so doing one is led 
to identify the classical action for which such paths are appropriate. If 
we proceed in the same fashion, we arrive at the formal functional 
integral expression
$$ {\cal M}\int e^{i\tint_0^T[p\cdot{\dot q}-H(p,q,t)-
\l^a(t)\phi_\a(p,q)]\,dt}\,\D p\,\D q\,\D R(\l)\;. $$
In this expression we have introduced the symbols   
\b &&H(p,q,t)\equiv\<p,q|\H(P,Q,t)|p,q\>\;,\\
&&\hskip.26cm\phi_\a(p,q)\equiv\<p,q|\Phi_\a(P,Q)|p,q\>\;.\ee 
We further note that $H(p,q,t)=\H(p,q,t)+O(\hbar^c)$, $c>0$, and 
$\phi_\a(p,q)=\Phi_\a(p,q)+O(\hbar^d)$, $d>0$, thus leading to the 
expected behavior as $\hbar\ra0$.   

It is clear from this expression that the system we have quantized here 
is determined by the ($\hbar$ augmented) classical 
action functional
$$\tint_0^T[p\!\cdot\!{\dot q}-H(p,q,t)-\l^\a(t)\,\phi_\a(p,q)]\,dt\;,  $$
which has the expected form of a canonical system in the presence of a 
number of constraints. Evoking a stationary action principle for this 
classical action leads to both dynamical and constraint equations as usual.
Note that the second-order terms in the constraint operators needed in 
the lattice formulation to construct the correct quantum projection 
operator disappear in this formal expression
since, for paths and Lagrange multiplier variables that are smooth and 
differentiable, i.e., have classical behavior, these second-order terms 
formally disappear.

While the derived classical action essentially reflects the original 
quantum theory, it is necessary, as always, to exercise great care in 
elevating a formal path integral involving the classical action to a 
valid calculational procedure. While not every lattice-limit formulation 
will lead to the correct answer, we emphasize that one acceptable 
construction procedure is the one presented in this paper. 
\subsubsection*{Commentary}
In a classical theory, such as the one above, it is true that the 
Lagrange multiplier variables play a significant role in the solution 
of the classical equations of motion, either being fixed functions 
chosen to enforce the 
constraints in 
the case of second-class constraints, 
or needing some arbitrary specification (selection of a gauge) 
when the Lagrange 
multiplier variables 
are not determined by the equations of motion in the 
case of first-class constraints. However, what holds true 
in the classical theory has no reason to hold true in the completely 
different quantum theory. For example,  
it should be noted that nowhere in our discussion have we used the 
words ``gauge'' or ``gauge fixing'' in achieving the proper quantum 
results. Indeed, in the abstract formulation of Section 2, the concepts 
of ``gauge'' and ``gauge fixing'' are completely foreign to the 
whole procedure. If anything, ``gauge fixing'', and its associated 
difficulties, 
is a classical concept inappropriately and unnecessarily carried 
over into the 
quantum theory. In concrete terms, and with respect to the formal 
phase space path integral we have derived, we note that no discussion of 
gauge 
invariance of the formal path integral has been given nor is any needed. 

If the constraints are all first class and closed, then gauge invariance of 
the 
resulting theory as expressed in the (true) physical Hilbert space may be 
demonstrated \cite{kla}. On the one hand, if $\d>0$ is chosen sufficiently 
small so that $\E(\Phi_\a M^{\a\beta}\Phi_\beta\le\d^2)\equiv
\E(\Phi_\a M^{\a\beta}\Phi_\beta=0)\ne0$, then invariance of the resulting 
theory in the physical Hilbert space follows from the fact that 
$e^{-i\omega^\a\Phi_\a}\,\E=\E$ for arbitrary real-valued 
$\{\omega^\a\}_{\a=1}^A$. On the other hand, if 
$\E(\Phi_\a M^{\a\beta}\Phi_\beta\le\d^2)\ne\E(\Phi_\a 
M^{\a\beta}\Phi_\beta=0)$ no matter how small $\d>0$ is, 
then it means that $e^{-i\omega^\a\Phi_\a}\,\E=\E+O(\d\!\cdot\!\E)$, 
and gauge invariance arises in the true physical Hilbert space 
${\frak H}_{\,\rm true\;phys}$
only after a suitable $\d\ra0$ limit as illustrated in Appendix B. 
Of course, for second-class constraints the issue of gauge invariance does
not arise. The presence of such constraints is distinguished by the fact 
that $\E\equiv0$ for all $\d<\d'$ for some $\d'>0$. 

Additionally, we observe that the abstract operator formalism 
developed in Section 2 applies to a field theory just as well as 
to general systems of finitely many degrees of freedom, canonical or 
otherwise. To discuss any system---a model field theory included---one 
only needs to have candidate operators for the Hamiltonian and for the 
various constraints expressed in the original Hilbert space.

\section*{Acknowledgements} The comments of B.G. Bodmann, J. Govaerts, 
and S.V. Shabanov are gratefully acknowledged.
\vfill\eject

\section*{Appendix A}
The construction of the first projection operator example in Section 1 
relies on a conditionally convergent integral, defined for 
$-\infty<x<\infty$, which is given by
$$ \int_{-\infty}^\infty\,e^{-ix\l}\,\frac{\sin(\d\l)}{\pi\l}\,d\l\;. $$
We may assign one meaning to this expression by choosing the formula
 $$ \lim_{L\ra\infty}\int_{-L}^L\,e^{-ix\l}\,
\frac{\sin(\d\l)}{\pi\l}\,d\l\;, $$
an expression which defines the function
 \b  f_{\frac{1}{2}}(x)&=& 1\;\;{\rm if}\;|x|<\d\;,\\
     &=& \half\;\;{\rm if}\;|x|=\d\;, \\
     &=& 0\;\;{\rm if}\;|x|>\d\;.  \ee
It is clear that $f_{\frac{1}{2}}(x)^2\not\equiv f_{\frac{1}{2}}(x)$, 
specifically at $|x|=\d$, which would be needed to ensure that $\E^2=\E$.

To overcome this difficulty, we adopt the prescription
 \b f_1(x)&\equiv&\lim_{\zeta\ra0^+}\lim_{L\ra\infty}\int^L_{-L}\,
e^{-ix\l}\,\frac{\sin[(\d+\zeta)\l]}{\pi\l}\,d\l\\
 &=&1\;\;{\rm if}\;|x|\le\d\;,\\
 &=&0\;\;{\rm if}\;|x|>\d\;.  \ee
Clearly in this case $f_1(x)^2\equiv f_1(x)$ for all $x$. Likewise we 
can also define 
\b f_0(x)&=&1\;\;{\rm if}\;|x|<\d\;,\\
   &=&0\;\;{\rm if}\;|x|\ge\d\;,  \ee
which also satisfies $f_0(x)^2\equiv f_0(x)$ for all $x$, by replacing 
$\lim_{\zeta\ra0^+}$ by $\lim_{\zeta\ra0^-}$ in the definition of $f_1(x)$.

In application to a single, self-adjoint quantum constraint $\Phi$, we 
conclude, for any $\d>0$, that 
$$ \E(\Phi^2\le\d^2)=\lim_{\zeta\ra0^+}\lim_{L\ra\infty}\int^L_{-L}\,
e^{-i\Phi\l}\,\frac{\sin[(\d+\zeta)\l]}{\pi\l}\,d\l $$
and
$$ \E(\Phi^2<\d^2)=\lim_{\zeta\ra0^-}\lim_{L\ra\infty}\int^L_{-L}\,
e^{-i\Phi\l}\,\frac{\sin[(\d+\zeta)\l]}{\pi\l}\,d\l\;. $$
In many cases the resulting projection operators are identical. An 
example for which they are not identical is given by 
$$|0\>\<0|\equiv\E(P^2+Q^2\le\hbar)\not=\E(P^2+Q^2<\hbar)\equiv0\;. $$
Here the one-dimensional projection operator $|0\>\<0|$ projects
onto the ground state for a harmonic 
oscillator ``Hamiltonian'' $P^2+Q^2$.

Unless stated otherwise, we shall adopt the convention that
 $$\int^\infty_{-\infty}\,e^{-i\Phi\l}\,\frac{\sin(\d\l)}{\pi\l}\,
d\l\equiv\E(\Phi^2\le\d^2)\;. $$

\section*{Appendix B}
The issue of dealing with constraint operators with their zero in the 
continuous spectrum has been dealt with by several authors \cite{hur}. 
Here we offer a brief discussion in keeping with the style of the 
present paper.
 
Let us focus on the projection operator $\E(X^2\le\d^2)$ where $X=0$ 
lies in the continuous spectrum of the self-adjoint operator $X$ 
defined in the original Hilbert space $\frak H$. We write 
${\frak H}={\frak H}_1\otimes{\frak H}_2$ such that, for sufficiently 
small $\d$, $X=X_1\otimes\one_2$, where $X_1$ is nondegenerate while 
${\frak H}_2$  accounts for possible degeneracy. Generally, this 
factorization depends on $\d$, but we assume that a suitable limit of 
these spaces exists as $\d\ra0$. In the same decomposition, 
$\E=\E_1\otimes\one_2$. For product vectors $|\psi\>=|\psi_1\>
\otimes|\psi_2\>\in{\frak H}$, it follows that
 \b &&\<\psi|\E(X^2\le\d^2)|\phi\>=\<\psi_1|\E_1(X_1^2\le\d^2)|
\phi_1\>\<\psi_2|\phi_2\>\\
  &&\hskip3.35cm =\int^\d_{-\d}\,d\<\psi_1|E_1(x)|\phi_1\>\;
\<\psi_2|\phi_2\>\;, \ee
where $E_1(x)$ is the spectral family for $X_1$.
For the first factor we choose a functional representation that 
diagonalizes $X_1$ leading to
$$ \int^\d_{-\d}\,\psi_1(x)^*\phi_1(x)\,dx \;.  $$
For every $|\psi_1\>$ and $|\phi_1\>$, this expression vanishes as 
$\d\ra0$. However, before taking the limit let us first rescale 
this integral so that it reads
 $$ \frac{1}{2\d}\,\int^\d_{-\d}\,\psi_1(x)^*\phi_1(x)\,dx \;. $$
This expression has a well-defined and generally nonvanishing limit 
for a {\it subset} of elements $\psi_1(x)$ and $\phi_1(x)$, namely, 
those that are continuous functions in an open set about $x=0$. For the 
subset of continuous functions, it follows that
 $$\lim_{\d\ra0}\frac{1}{2\d}\,\int^\d_{-\d}\,\psi_1(x)^*\phi_1(x)\,dx=
\psi_1(0)^*\phi_1(0)\;. $$
To summarize, we determine, for suitable elements $|\psi_1\>$ and 
$|\phi_1\>$, that 
$$ \lim_{\d\ra0}\,(2\d)^{-1}\<\psi|\E(X^2\le\d^2)|\phi\>=
\psi_1(0)^*\phi_1(0)\,\<\psi_2|\phi_2\>\;. $$
If this expression is regarded as an incomplete inner product, then, 
after completion, we learn that the resultant space is ${\mathbb C}
\otimes{\frak H}_2\simeq{\frak H}_2$ since $\mathbb C$, the space
of complex numbers, has only one (complex)
dimension. We conclude that ${\frak H}_{\,\rm true\; phys}={\frak H}_2$. 

We can understand this result in more physical terms. Let us realize 
${\frak H}_1$ as $L^2([-\d,\d])$ and introduce an orthonormal basis 
given by $u_n(x)=(1/\sqrt{2\d})\,e^{in\pi x/\d}$, 
$n\in\{0,\pm 1,\pm2,\dots\}$. If the Hamiltonian contained the term 
$P_x^2=-\partial^2/\partial x^2$, or such a term was added to the 
Hamiltonian, then 
$$ P_x^2\,u_n(x)=(n\pi/\d)^2\,u_n(x) $$
since $u_n$ is an eigenfunction of $P_x^2$. For very small $\d$, all 
energy contributions are huge save for the single case $n=0$ when 
this contribution 
vanishes. Thus, when $\d\ll1$ energetics favor only the ground state 
$u_0(x)=1/\sqrt{2\d}$ in an expansion of $\psi_1(x)$ and $\phi_1(x)$, 
and consequently
\b &&\lim_{\d\ra0}\,\frac{1}{2\d}\int^\d_{-\d}\psi_1(x)^*\phi_1(x)\,dx=
\lim_{\d\ra0}\frac{1}{(2\d)^2}\int_{-\d}^\d\psi_1(y)^*\,dy\,
\int^\d_{-\d}\phi_1(z)\,dz\\
  &&\hskip4.75cm =\psi_1(0)^*\phi_1(0)  \ee
for functions which are continuous about $x=0$.

Is it really necessary to take the limit $\d\ra0$? In our view the 
practical answer is ``No''. For example, if energies are restricted to 
those which are much less than $\d^{-2}$, a number which can be chosen 
as large as one likes, then only the ground-state contribution arises 
for ${\frak H}_1$ and hence ${\frak H}_{\,\rm true \;phys}=
{\frak H}_{\,\rm phys}$, as desired.
\vfill\eject

\end{document}